\documentclass[12pt]{article}

\usepackage{mathrsfs}
\usepackage{amsmath}

\global\arraycolsep=1pt
\usepackage[colorlinks=true,backref=true,linkcolor=black,anchorcolor=black,citecolor=black,filecolor=black,menucolor=black,pagecolor=black,urlcolor=black]{hyperref}
\usepackage[Symbol]{upgreek}
\usepackage{bbm}
\usepackage{dsfont}
\usepackage{amssymb}
\usepackage{textcomp}

\newcommand{\Scal}[1]{\Bigl ({#1} \Bigr )}
\newcommand{\scal}[1]{\bigl ({#1} \bigr )}

\newcommand{\CR}{\nonumber \\*}

\newcommand{\trace}{\hbox {Tr}~}

\newcommand{\gra}[2]{{\scriptscriptstyle (#1 , #2 )}}

\DeclareMathAlphabet{\mathpzc}{OT1}{pzc}{m}{it}
\def\s{\,\mathpzc{s}\,}

\def\d{\mathfrak{d}}

\usepackage[colorlinks=true,backref=true,linkcolor=black,anchorcolor=black,citecolor=black,filecolor=black,menucolor=black,pagecolor=black,urlcolor=black]{hyperref}

\usepackage[Symbol]{upgreek}
\usepackage{caption}
\usepackage{bbm}
\usepackage{dsfont}
\usepackage{amssymb}
\usepackage{textcomp}

\def\N{\mathcal{N}}

\def\d{\delta}

\def\ga{\gamma}
\def\te{\theta}
\def\vte{\vartheta}

\def\tga{\tilde{\gamma}}

\def\be{\begin{equation}}
\def\ee{\end{equation}}
\def\bea{\begin{eqnarray}}
\def\eea{\end{eqnarray}}
\def\bdis{\begin{displaymath}}
\def\edis{\end{displaymath}}

\def\d{\delta}

\def\st{{\theta}}

\begin{document}
\allowdisplaybreaks[1]
\renewcommand{\thefootnote}{\fnsymbol{footnote}}
\def\corr{$\spadesuit $}
\def\trefle{$\clubsuit$}

\renewcommand{\thefootnote}{\arabic{footnote}}
\setcounter{footnote}{0}


 \def\stop{$\blacksquare$}
\begin{titlepage}
\renewcommand{\thefootnote}{\fnsymbol{footnote}}
\begin{flushright}
\
\vskip -3cm
{\small CERN-PH-TH/2008-029}\\
\vskip 3cm
\end{flushright}
\begin{center}
{{\Large \bf

Twisted Superspace

 }}
\lineskip .75em
\vskip 3em
\normalsize
{\large Laurent Baulieu\footnote{email address: baulieu@lpthe.jussieu.fr},
 Guillaume Bossard\footnote{email address: bossard@aei.mpg.de},
Alexis Martin\footnote{email address: alexis.martin@lpthe.jussieu.fr}\\
\vskip 1em
$^{* }${\it Theoretical Division CERN} \footnote{ CH-1211 Gen\`eve, 23, Switzerland}
\\
$^{* \ddagger}${\it LPTHE, CNRS and Universit\'e Pierre et Marie
Curie } \footnote{ 4 place Jussieu, F-75252 Paris Cedex 05,
France}
\\
$^{\dagger}${\it AEI, Max-Planck-Institut f\"{u}r Gravitationsphysik} \footnote{Am M\"{u}hlenberg 1, D-14476 Potsdam, Germany}
}

\vskip 1 em
\end{center}
\vskip 1 em
\begin{abstract}
We formulate the ten-dimensional super-Yang--Mills theory in a twisted superspace with $8+1$ supercharges. Its constraints do not imply the equations of motion and we solve them. As a preliminary step for a complete formulation in a twisted superspace, we give a superspace path-integral formulation of the $\N=2,\ d=4$ super-Yang--Mills theory without matter. The action is the sum of a Chern--Simons term depending on a super-connection plus a $BF$-like term. The integration over the superfield $B$ implements the twisted superspace constraints on the super-gauge field, and the Chern--Simons action reduces to the known action in components.
\end{abstract}

\end{titlepage}
\renewcommand{\thefootnote}{\arabic{footnote}}
\setcounter{footnote}{0}

\section{Introduction}

Superspace formulations of supersymmetric theories are often very
efficient for practical computations and proofs of
non-renormalization theorems. A complete superspace
path-integral formulation requires that the supersymmetry algebra
admits a functional representation on the fields, but the latter
is believed not to exist for maximal supersymmetry. This has lead
to several proposals for restricting the whole super-Poincar\'e
algebra to subalgebras with such a representation. For instance, the maximally supersymmetric Yang--Mills
theory has been formulated within $\N = 3$ harmonic superspace
\cite{N=3}. There are severe restrictions on such
off-shell closed representations. In six dimensions,    to  maintain  the full
Poincar\'e invariance, one must reduce
  the $\N=2$ super-Poincar\'e   symmetry to the $\N=1$ one. From
dimension seven and above, no non-trivial subalgebra includes
the whole Poincar\'e algebra. In fact, a superspace
path-integral formulation of maximally supersymmetric Yang--Mills
theories in higher dimensions must presumably give up manifest
Lorentz invariance.

 In \cite{10DSYM} we have shown that $SO(1,1)\times Spin(7) \subset SO(1,9) $ is the biggest subgroup of the ten-dimensional Lorentz group that can be preserved for obtaining an off-shell closed supersymmetric algebra of the $\N=1,d=10$ supersymmetric theory. We introduced for this theory $SO(1,1)\times Spin(7) \subset SO(1,9) $ invariant constraints for the curvatures of superfields depending of $1+8$ fermionic coordinates, as a hint for a possible off-shell superspace description.
   Part of this Letter is devoted to solve explicitly these constraints, in function of the fields of the  component formalism, in dimension up to $d=10$.

The maximally supersymmetric theory in ten dimensions is a chiral
model with a gauge anomaly that spoils its quantization. A
consistent approach implies in fact its coupling to supergravity
at the quantum level. However, its chiral anomaly often disappears
after dimensional reduction to lower dimensions.
It is thus a relevant question to investigate a possible
superspace off-shell formulation of the pure ten-dimensional
supersymmetric Yang--Mills theory.

The path integral quantization procedure in superspace usually
requires to solve the superspace constraints by introduction of an
unconstrained prepotential. This procedure is difficult  within 
 the considered twisted superspace, mainly because
some constraints are cubic in the gauge superfields. This justifies that 
 we first consider the quantization of a simpler model in
four dimensions. As a matter of fact, the twisted $SU(2)\times
SU(2)$ invariant formulation of the $\N=2$ super-Yang--Mills
theory without matter in four dimensions is formally very similar
to the discussed formulation of the ten-dimensional theory. 
Instead of introducing a prepotential, we quantize this  theory 
by implementing the constraints by mean of Lagrange
multipliers. The path-integral is then formulated in term of the unconstrained potential superfields themselves.


The  four-dimensional  action is written as an integral over the full twisted superspace of three different parts. The superspace constraints are implemented by Lagrange multipliers through a $BF$-like action.
 where $F$ stands for the components of the super-curvature that define the constraints.
Because of Bianchi identities,  the auxiliary superfields $B$ possess a set of zero modes that must be taken into account in the super-Feynman rules. The complete gauge-fixing of the $BF$ component of the action requires the introduction of an infinite tower of ghosts and ghosts for ghosts. This problem is reminiscent of the infinite set of auxiliary fields required in the harmonic-superspace formulation of the theory \cite{harmonic}. 
The classical part is a Chern--Simons-like action for the superspace connection along the scalar odd coordinate. 
Finally, the gauge-fixing part is a generalization in superspace of the usual Landau gauge-fixing action. Its decomposition in components turns out to be equivalent to a supersymmetric gauge-fixing involving shadow fields, which generalize those introduced in \cite{shadow}.

 We will
ignore through the Letter the problems associated with unitarity and the doubling of fermions in four and eight-dimensional euclidean space. This is justified within the context
 of describing the ten-dimensional structure.


%


The Letter is organized as follows. In the first section we define the $\N=2$ twisted superspace, and its generalization in higher dimensions. Then we define the twisted super-Yang--Mills constraints and solve  them in term of the component fields in four dimensions and generalize the results in ten dimensions, with an obvious application in eight dimensions. In the last section, we construct the action for the $\N=2, d=4$ theory.  We explain the problem associated with the gauge invariance of the Lagrange multipliers that enforce the covariant constraints on the supercurvature, but postpone to a forthcoming publication the definition of the corresponding gauge-fixing action.

There are earlier references  for  the idea of  twisted
superspace  for the $\N=2, d=4$ twisted
super-Yang--Mills vector multiplet  
\cite{TopolMatter,N4twisted,N2hyper}.
The general  superspace    methodology for super-connections    relies on~\cite{Sohnius,Sohnius&al}.
 No path-integral formulation in twisted
superspace had  been proposed so far.

\section{Twisted superspace set-up}

\subsection{ The $\N=2$, $d=4$ case}

Let us recall the basic features of the twisted formulation of $\N=2$
super-Yang--Mills theory \cite{AlgRenorm}. It is defined in a four-dimensional euclidean space
with the manifest invariance reduced to $L^\prime= SU(2)^\prime \times SU(2)_R$,
where $SU(2)^\prime$ is the diagonal subgroup of $SU(2)_L \times
SU(2)_I$, and $SU(2)_I$ is the internal symmetry group associated
to $\N=2$ supersymmetry. The vector multiplet in representations of $L^\prime$ is made of the
gauge field $A_\mu$, two commuting scalar fields $\Phi $ and
$\bar\Phi$, an anticommuting vector $\Psi_\mu$, an anticommuting
anti-selfdual 2-form $\chi_{\mu\nu^-}$, an anticommuting scalar
$\eta$, and a commuting auxiliary field $G_{\mu\nu^-}$.
These fields transform under the scalar and vector anticommuting
generators $\d\equiv\epsilon^{\alpha \imath}Q_{\alpha \imath}$ and
$\d_\mu\equiv i \sigma_\mu^{\dot{\alpha} \imath}Q_{\dot{\alpha} \imath}$.
The invariance under the action of these 5 generators completely
determines the classical action of the theory, which is nothing but the super-Yang--Mills action, in twisted form \cite{shadow}. In order to recover
the complete super-Poincar\'e symmetry with 8 generators, one must
introduce the anti-selfdual generator $\d_{\mu\nu^-} \equiv
\sigma_{\mu\nu}^{\alpha \imath} Q_{\alpha \imath}$. The $\d_{\mu\nu^-}$
invariance is an additional symmetry of the action, which is
obtained for free from the requirement of $\d$ and $\d_\mu$
symmetry. Moreover the absence of trivial anomalies for the tensor symmetry
shows that
forgetting about the tensor symmetry does not introduce
ambiguities in the renormalization program \cite{TBP}. Therefore, as long as
we only consider correlation functions of the fields, the scalar
and vector supersymmetry generators unambiguously determine the
theory to be invariant by the action of all the supersymmetry
generators, including the tensor generator $\d_{\mu\nu^-}$.

To express the scalar and vector supersymmetry in terms of superspace
derivatives, we complete the four-dimensional space by five
anticommuting coordinates, a scalar one $\te$ and a vector one
$\vte^\mu$ ($\mu=1\cdots 4$). We define as follows the superspace
differential operators $\mathbb{Q}$ and $\mathbb{Q}_\mu$, whose
action on superfields provide  component by component a linear realization
of the scalar $\delta$ and vector $\delta_\mu$  supersymmetry  generators
\bea \label{susy gen} \mathbb{Q}\,\, \equiv
\frac{\partial}{\partial\te}+\vte^\mu\partial_\mu, \hspace{20mm}
\mathbb{Q}_\mu \equiv\frac{\partial}{\partial\vte^\mu} \CR
\mathbb{Q}^2=0,\hspace{10mm} \{\mathbb{Q},\mathbb{Q}_\mu\}=
\partial_\mu, \hspace{10mm} \{\mathbb{Q}_\mu, \mathbb{Q}_\nu\}=0
\eea
A general superfield $\mathbb{S}_A$ is a polynomial expansion in ($\te,\vte^\mu$)
\be \mathbb{S}_A =\mathbf{S}^0_A + \te\mathbf{S}^\te_A = S_A + \vte^\mu S_{A\mu}+\vte^\mu \vte^\nu
S_{A\mu\nu}+\cdots +\te S^\te_A +\te\vte^\mu S^\te_{A\mu}+\cdots
\ee
Here the index $A$ stands for the $L^\prime$ representation of the superfield and
$\mathbb{S}_A $ carries $\sharp(A)\times 2^5$ components,
where $\sharp(A)$ is the dimension of the corresponding $L^\prime$ representation.

The covariant superspace derivatives and their anticommuting relations are
\bea \nabla\,\, \equiv
\frac{\partial}{\partial\te} \hspace{20mm} \nabla_\mu
\equiv\frac{\partial}{\partial\vte^\mu}-\te\partial_\mu \CR
\nabla^2=0 \hspace{10mm} \{\nabla,\nabla_\mu\}= -\partial_\mu
\hspace{10mm} \{\nabla_\mu, \nabla_\nu\}=0 \eea They anticommute with the supersymmetry generators.

A connection superfield
$(\mathbb{C},\mathbf{\Upgamma}_\mu,\mathbb{A}_\mu)$ valued in the
adjoint of the gauge group of the theory can be defined in
correspondence with the set of the superspace derivatives
$(\nabla,\nabla_\mu,\partial_\mu)$. This provides the following
gauge covariant superderivatives \be \hat{\nabla} \equiv \nabla +
\mathbb{C},\quad \hat{\nabla}_\mu \equiv
\nabla_\mu+\mathbf{\Upgamma}_\mu,\quad \hat{\partial}_\mu\equiv
\partial_\mu+\mathbb{A}_\mu \ee The corresponding covariant
superspace curvatures   are    \be \label{ricci}\begin{split}
&\mathbb{F}_{\mu\nu} \equiv [\hat{\partial}_\mu,\hat{\partial}_\nu] \\
&\mathbf{\Uppsi}_\mu \equiv [\hat{\nabla},\hat{\partial}_\mu] \\
&\boldsymbol{\upchi}_{\mu\nu} \equiv
[\hat{\nabla}_\mu,\hat{\partial}_\nu]
\end{split}
\hspace{10mm}\begin{split}
&\mathbf{\Upphi} \equiv \hat{\nabla}^2 \\
&\mathbb{L}_\mu \equiv \{\hat{\nabla}, \hat{\nabla}_\mu\}+\hat{\partial}_\mu \\
&\mathbf{\bar{\Upphi}}_{\mu\nu} \equiv
{\ \scriptstyle \frac{1}{2} } \{\hat{\nabla}_\mu,\hat{\nabla}_\nu \}
\end{split}\ee
so that \be \label{curv}\begin{split}
\mathbb{F}_{\mu\nu} &= \partial_\mu\mathbb{A}_\nu -\partial_\nu\mathbb{A}_\mu +[\mathbb{A}_\mu,\mathbb{A}_\nu] \\
\mathbf{\Uppsi}_\mu &= \nabla \mathbb{A}_\mu - \partial_\mu\mathbb{C} - [\mathbb{A}_\mu,\mathbb{C}] \\
\boldsymbol{\upchi}_{\mu\nu} &= \nabla_\mu\mathbb{A}_\nu - \partial_\nu \mathbf{\Upgamma}_\mu
- [\mathbb{A}_\nu,\mathbf{\Upgamma}_\mu]
\end{split}
\hspace{10mm}\begin{split}
\mathbf{\Upphi} &=\nabla\mathbb{C}+\mathbb{C}^2 \\
\mathbb{L}_\mu &= \nabla\mathbf{\Upgamma}_\mu +\nabla_\mu\mathbb{C}+\{\mathbf{\Upgamma}_\mu, \mathbb{C}\}+\mathbb{A}_\mu \\
\mathbf{\bar{\Upphi}}_{\mu\nu} &= \nabla_{\{ \mu} \mathbf{\Upgamma}_{\nu\}}
+ \mathbf{\Upgamma}_{\{\mu} \mathbf{\Upgamma}_{\nu\}}
\end{split}\ee
These different objects can be assembled into an extended exterior
differential
 \be \mathbf{\Updelta} \equiv d+\nabla d\st +\nabla_{d
 \vte } \equiv
 dx^\mu \partial_\mu + d \st \nabla + {d\vte}^\mu \nabla_\mu
\ee
and the extended connection
 \be \mathcal{A}\equiv
\mathbb{A}+\mathbb{C} d\st +\mathbf{\Upgamma} \equiv \mathbb{A}_\mu
dx^\mu+\mathbb{C}d\st +\mathbf{\Upgamma}_\mu d\vte^\mu\ee
Since $(d+\nabla d\st +\nabla_{d \vte }+ d\st i_{d \vte
}) ^2=0$, (where $ i $ is the Cartan contraction operator, e.g., $ i_{d \vte } dx^\mu\equiv {d \vte } ^\mu$), we    define the following extended   curvature superfield $2$-form $\mathcal{F}$
 \bea\label{curvature}
 \mathcal{F}\equiv
(d+\nabla d\st +\nabla_{d \vte }+ d\st i_{d \vte
}) \mathcal{A}+ \mathcal{A}^2
=
\mathbb{F}+\mathbf{\Uppsi}d\st +\boldsymbol{\upchi}+\mathbf{\Upphi}d\st d\st +\mathbb{L}d\st +\mathbf{\bar{\Upphi}}
\eea
 where $
 \mathbb{F}\equiv \frac{1}{2}\mathbb{F}_{\mu\nu}dx^\mu
dx^\nu,
\ \mathbf{\Uppsi}\equiv \mathbf{\Uppsi}_\mu dx^\mu ,\
\boldsymbol{\upchi}\equiv \boldsymbol{\upchi}_{\mu\nu} d\vte^\mu
dx^\nu,
\ \mathbb{L}\equiv \mathbb{L}_\mu d\vte^\mu ,\
\mathbf{\bar{\Upphi}}\equiv \mathbf{\bar{\Upphi}}_{\mu\nu} d\vte^\mu
d\vte^\nu$.
The Bianchi identity implies the following constraints on the components of $ \mathcal{F}$
 \begin{multline} (d+ d\st \nabla+\nabla_{d
\vte }+ d\st i_{d \vte
})(\mathbb{F}+\mathbf{\Uppsi}d\st +\boldsymbol{\upchi}+\mathbf{\Upphi}d\st d\st +\mathbb{L}d\st +\mathbf{\bar{\Upphi}})+\\
[ \mathcal{A},\mathbb{F}+\mathbf{\Uppsi}d\st+\boldsymbol{\upchi}+\mathbf{\Upphi}d\st d\st+\mathbb{L}d\st+\mathbf{\bar{\Upphi}}]
= 0 \end{multline}
The  super-gauge transformations of the  extended connection $\mathcal{A} $ and   curvature $\mathcal{F} $  are  \be
\label{gauge_transf} \mathcal{A} \rightarrow
e^{-\boldsymbol{\upalpha}}(\mathbf{\Updelta}+\mathcal{A})e^{\boldsymbol{\upalpha}},
\quad \mathcal{F} \rightarrow
e^{-\boldsymbol{\upalpha}}\mathcal{F}e^{\boldsymbol{\upalpha}} \ee
where the gauge superparameter $\boldsymbol{\upalpha}$ can be any
given general superfield valued in the Lie algebra of the gauge
group.  The  ``infinitesimal" gauge transformation is
$\delta \mathcal{A} =\mathbf{\Updelta} \boldsymbol{\upalpha} +[\mathcal{A} ,\boldsymbol{\upalpha} ]$.

 \subsection{Higher dimensions}

The formalism for the scalar and vector supersymmetry generalizes directly to the euclidean eight-dimensional case, by extending the eight-dimensional space-time with nine fermionic coordinates and considering a reduction of the Wick rotated Lorentz group $SO(8)$ to $Spin(7)$, with all previous equations remaining formally identical. One  can    further  ``oxidise"  the   eight-dimensional theory  into the $\N=1, d=10 $ theory. This  has already been described in \cite{10DSYM}, and we shall only  summarise the equations that are relevant for the following.  (One can  go  from  four to six dimensions in an analogous way).

The $\N=1$, $d=10$ superspace is made of ten bosonic coordinates $x^m$ and nine fermionic ones $\te$ and $\vte^i$. The $x^m$ ($m=0,\cdots 9$) split into euclidean eight-dimensional coordinates $x^i$ and light-cone coordinates $x^+$ and $x^-$, so that a general ten-dimensional form splits as $\mathbb{A}_m dx^m=\mathbb{A}_i dx^i +\mathbb{A}_+ dx^+ +\mathbb{A}_- dx^-$. The Grassmann  coordinates  $\te$ and $\vte^i$   are  scalar and   vector,  the latter being identified with  the spinorial representation $\mathbf{8}$ of $Spin(7)$. The covariant superspace derivatives are defined as $\nabla\,\,\equiv \frac{\partial}{\partial\te}-\te\partial_+$ and $\nabla_i
\equiv\frac{\partial}{\partial\vte^i}-\te\partial_i-\vte_i\partial_-$, with
\be
\nabla^2=-\partial_+,\hspace{10mm} \{\nabla,\nabla_i\}= -\partial_i,
\hspace{10mm} \nabla_{\{i} \nabla_{j\}}=-\delta_{ij}\partial_-
\ee
Super-curvatures are defined by the analogue of Eq.(\ref{curvature}) for ten dimensions
\bea\label{curvature10D}
(d+d\st \nabla+\nabla_{d {\vte} }+  i_{(d\st^2\partial_+ +d\st d \vte + \vert d\vte\vert^2\partial_-
)})(\mathbb{A}+\mathbb{C}d\st +\mathbf{\Upgamma})+(\mathbb{A}+\mathbb{C}d\st +\mathbf{\Upgamma})^2\CR
=
\mathbb{F}+\mathbf{\Uppsi}d\st +\boldsymbol{\upchi}+\mathbf{\Upphi}d\st d\st +\mathbb{L}d\st +\mathbf{\bar{\Upphi}}
\eea
where $\mathbb{F}\equiv \frac{1}{2}\mathbb{F}_{mn}dx^m
dx^n, \mathbf{\Uppsi}\equiv \mathbf{\Uppsi}_m dx^m,
\boldsymbol{\upchi}\equiv \boldsymbol{\upchi}_{in} d\vte^i
dx^n, \mathbb{L}\equiv \mathbb{L}_i d\vte^i$ and
$\mathbf{\bar{\Upphi}}\equiv \mathbf{\bar{\Upphi}}_{ij} d\vte^i
d\vte^j$. One has in particular\footnote{We have analogous notations $\mathbf{\Upphi}$ and $\bar{\mathbf{\Upphi}}_{\alpha\beta}$  for the curvatures of the different  $\N=1$ and $\N=2$ cases, in six (respectively four) dimensions ($\alpha,\beta \hat{=} \mu,\nu$), and ten (respectively eight) dimensions ($\alpha,\beta \hat{=} i,j$). However, after dimensional reduction and once the constraints $\mathbf{\Upphi}^{ \N=1}=\bar{\mathbf{\Upphi}}_{ij}^{ \N=1}=0$ are imposed, we have the correspondence $\mathbb{A}_+ \rightarrow \mathbf{\Upphi}^{\N=2}$ and $\mathbb{A}_-\rightarrow \bar{\mathbf{\Upphi}}^{\N=2}$.}
\be
\mathbf{\Upphi}\equiv\hat{\nabla}^2+\hat{\partial}_+,\hspace{10mm} \mathbb{L}_i \equiv \{\hat{\nabla}, \hat{\nabla}_i\}+\hat{\partial_i},\hspace{10mm} \mathbf{\bar{\Upphi}}_{ij} \equiv
\hat{\nabla}_{\{i} \hat{\nabla}_{j \}}+ \d_{i j}\hat{\partial_-} .
\ee

\section{Constraints and their resolution}

\subsection{ The $\N=2$, $d=4$ case}
To eliminate superfluous degrees of freedom and to make contact
with the component formulation, we must impose
superspace gauge covariant constraints, as follows  \be
\label{constraints} \mathbb{L}_\mu = 0,\quad
\bar{\mathbf{\Upphi}}_{\mu\nu}=\frac{1}{4}\delta_{\mu\nu}\bar{\mathbf{\Upphi}}_\sigma^{\phantom{\sigma}\sigma}\equiv\delta_{\mu\nu}\bar{\mathbf{\Upphi}},
\quad \boldsymbol{\upchi}_{[\mu\nu]_+}=0. \ee The super-gauge
symmetry defined in Eq.(\ref{gauge_transf}) allows us to simplify
the resolution of the constraints.
 We partially fix super-gauge invariance by setting  to zero   all antisymmetric components $(\frac{\partial}{\partial\vte^{[\mu}}\cdots\frac{\partial}{\partial\vte^\sigma}\boldsymbol{\Upgamma}_{\rho]})|_0$ and $(\frac{\partial}{\partial \theta} \frac{\partial}{\partial\vte^{[\mu}}\cdots\frac{\partial}{\partial\vte^\sigma}\boldsymbol{\Upgamma}_{\rho]})|_0$ of $\mathbf{\Upgamma}_\mu$, including  $\boldsymbol{\Upgamma}_\mu |_0$, as well as
 the first component $\mathbb{C}|_0$ of $\mathbb{C}$.\footnote{We use the standard notation $|_0$ for expressing that all fermionic coordinates are set to zero.} In this gauge, the remaining gauge invariance reduces to that
 of the component formalism ($\boldsymbol{\upalpha} = \boldsymbol{\upalpha}|_0$).
 The details of the procedure will be found in \cite{TBP}. After solving the constraints in this particular super-gauge, we will reintroduce the super-gauge invariance by a general gauge transformation depending on new fields that stand for the longitudinal components.

We start with $\mathbf{\Upgamma}_\mu$. The constraint
Eq.(\ref{constraints}) on $\bar{\mathbf{\Upphi}}_{\mu\nu}$
and its Bianchi identity leave its $\vte^\mu$
independent trace components unconstrained. We define them as
$\bar{\mathbf{\Upphi}}|_0
\equiv \bar{\Phi}$ and
$(\frac{\partial}{\partial\te}\bar{\mathbf{\Upphi}})|_0
\equiv \eta$. Using the definition of
$\bar{\mathbf{\Upphi}}_{\mu\nu}$ in terms of
$\mathbf{\Upgamma}_\mu$ and its Bianchi identity, we then obtain
 \be \mathbf{\Upgamma}_\mu =
\vte_{\mu}\bar{\Phi}+\te(\vte_\mu\eta+\vte_\mu\vte^\rho\partial_\rho\bar{\Phi}),\quad
\bar{\mathbf{\Upphi}}=\bar{\Phi}+\te(\eta-\vte^{\mu}\partial_{\mu}\bar{\Phi})
\ee
The constraint
$\mathbb{L}_\mu = 0$ allows us
to express $\mathbb{A}_\mu$ in terms of
$\mathbf{\Upgamma}_\mu$ and $\mathbb{C}$. It is convenient to parametrize the
 superfield $\mathbb{C}$ as \be \mathbb{C} \equiv
\tilde{A}+\te(\tilde{\Phi}-\tilde{A}^2) \quad\rightarrow\quad
\mathbf{\Upphi} = \tilde{\Phi}+\te [\tilde{\Phi},\tilde{A}] \ee
where $\tilde{\Phi}$ and $\tilde{A}$ are general functions in
$\vte$ variables, except that $\tilde{A}|_0=0$ as it is required by our
special gauge choice. Moreover, we define
$(\frac{\partial}{\partial\vte^\mu}\tilde{A})|_0\equiv A_\mu$ and
$\tilde{\Phi}|_0\equiv \Phi$. We can then determine
$\mathbb{A}_\mu$ as \be \mathbb{A}_\mu =
\frac{\partial}{\partial\vte^\mu}\tilde{A} + \cdots -\te \Scal{
\frac{\partial}{\partial\vte^\mu}\tilde{\Phi}+\cdots} \ee The
explicit content of $\tilde{\Phi}$ and $\tilde{A}$ is determined
through the resolution of the anti-selfdual constraint on the
$\boldsymbol{\upchi}_{\mu\nu}$ curvature. We first observe that the Bianchi identities and the constraint
$\mathbb{L}_\mu = 0$ imply \be
\boldsymbol{\upchi}_{\mu\nu}=-\delta_{\mu\nu}\scal{\nabla\bar{\mathbf{\Upphi}}+[\mathbb{C},\bar{\mathbf{\Upphi}}]}+\boldsymbol{\upchi}_{[\mu\nu]}\equiv
-\delta_{\mu\nu}\boldsymbol{\upeta}+\boldsymbol{\upchi}_{[\mu\nu]}
\ee This allows one to express $\boldsymbol{\upeta}$ and
$\boldsymbol{\upchi}_{[\mu\nu]}$ in terms of $\tilde{\Phi}$,
$\tilde{A}$ and $\bar{\Phi}$ and $\eta$,
\bea \boldsymbol{\upeta} &=& \eta + \vartheta^\mu \partial_\mu \bar \Phi + [ \tilde A , \bar \Phi ] + \cdots \CR
\boldsymbol{\upchi}_{[\mu\nu]} &=&
\frac{\partial}{\partial\vte^\mu}\frac{\partial}{\partial\vte^\nu}\tilde{A}+\cdots
+\te\left(\frac{\partial}{\partial\vte^\mu}\frac{\partial}{\partial\vte^\nu}\tilde{\Phi}
+\cdots\right) \eea The component $(\frac{\partial}{\partial
\vartheta^\mu} \tilde{\Phi})|_0$ is not constrained. We
define $(\frac{\partial}{\partial\vte^\mu}\tilde{\Phi})|_0\equiv
-\Psi_\mu$ and we solve the constraint
$\boldsymbol{\upchi}_{[\mu\nu]_+}=0$, component by
component. From the $\te$-independent part, we get
\be \tilde{A}
= \vte^\mu A_\mu
-\frac{1}{2}\vte^\mu\vte^\nu\chi_{\mu\nu}+\frac{1}{3!}\vte^\mu\vte^\nu\vte^\rho
\epsilon_{\mu\nu\rho}^{\phantom{\mu\nu\rho}\sigma}D_\sigma\bar{\Phi}
-\frac{1}{4!}\vte^\mu\vte^\nu\vte^\rho\vte^\sigma
\epsilon_{\mu\nu\rho\sigma}[\bar{\Phi},\eta] \ee
and the part proportional on $\te$ gives us that
\bea\label{tildephi} \tilde{\Phi}=\Phi -\vte^\mu\Psi_\mu
-\frac{1}{2}\vte^\mu\vte^\nu( F_{\mu\nu}+G_{\mu\nu})&+&
\frac{1}{3!}\vte^\mu\vte^\nu\vte^\rho\scal{3 D_\mu\chi_{\nu\rho}
- \epsilon_{\mu\nu\rho}^{\phantom{\mu\nu\rho}\sigma}(D_\sigma\eta
- [\bar{\Phi},\Psi_\sigma])} \CR -\frac{1}{4!}
\vte^\mu\vte^\nu\vte^\rho\vte^\sigma\left(2\epsilon_{\mu\nu\rho\sigma}
D_\lambda D^\lambda
\bar{\Phi} \right . &-& \left . 6\chi_{\mu\nu}\chi_{\rho\sigma}+2\epsilon_{\mu\nu\rho\sigma}\eta^2-
\epsilon_{\mu\nu\rho\sigma}[\bar{\Phi},[\bar{\Phi},\Phi]]\right)
\eea where $\chi, G$ are anti-selfdual $2$-forms and $F=dA+A^2$. As a result, the general solution of the constrained superfields in the chosen Wess--Zumino-like gauge can be written in term of the known component fields of the theory, with the auxiliary field required for the functional representation of the supersymmetry algebra.

The general solution to the constraints (\ref{constraints}) can now be obtained by application of a general super-gauge transformation, which we parametrize as follows\footnote{The gauge transformation is chosen in such a way as to recover the  transformation laws computed in components.}

\be
 e^{\boldsymbol{\upalpha}}=e^{\te\vte^\mu\partial_\mu}e^{\tilde{\gamma}}e^{\te\tilde{c}}=e^{\tilde{\gamma}}
\scal{1+\te( \tilde{c}+e^{-\tilde{\gamma}}\vte^\mu\partial_\mu
e^{\tilde{\gamma}})} \ee where $\tilde{\ga}$ and $\tilde{c}$ are
respectively commuting and anticommuting functions of $\vte^\mu$
and the coordinates $x^\mu$, with the condition
$\tilde{\ga}\vert_0=0$. The superfield connections $\mathbb{C}$,
$\mathbf{\Upgamma}$ and their curvatures then have the following
expressions \bea \mathbb{C} &=& \tilde{c} +
e^{-\tga}\left(\vte^\mu\partial_\mu +\tilde{A}\right)e^{\tga}
+\te\left(e^{-\tga}\tilde{\Phi}e^{\tga}-\left(\tilde{c}+e^{-\tga}\left(\vte^\mu\partial_\mu
+\tilde{A}\right)e^{\tga}\right)^2\right)\CR \mathbf{\Upphi} &=&
e^{-\tga}\tilde{\Phi} e^{\tga} +
\te\left(\left[e^{-\tga}\tilde{\Phi}e^{\tga},\tilde{c}\right]+e^{-\tga}\left[\tilde{\Phi},\vte^\mu\partial_\mu+\tilde{A}\right]e^{\tga}\right)\CR
\mathbf{\Upgamma}_\mu &=& e^{-\tga}\left(\frac{\partial}{\partial\vte^\mu}+\vte_\mu\bar{\Phi}\right)e^{\tga}+\te\bigg(e^{-\tga}\left(\vte_\mu\eta+\vte_\mu\vte^\rho\partial_\rho\bar{\Phi}\right)e^{\tga}\CR &&\hspace{50mm} \left.-\left[e^{-\tga}\left(\frac{\partial}{\partial\vte^\mu}+\vte_\mu\bar{\Phi}\right)e^{\tga},\tilde{c}+e^{-\tga}\vte^\mu\partial_\mu
e^{\tga}\right]\right)\CR \bar{\mathbf{\Upphi}} &=&
e^{-\tga}\bar{\Phi} e^{\tga} +
\te\left(e^{-\tga}\left(\eta-\vte^\mu\partial_\mu\bar{\Phi}\right)e^{\tga}+\left[e^{-\tga}\bar{\Phi}e^{\tga},\tilde{c}+e^{-\tga}\vte^\mu\partial_\mu
e^{\tga}\right]\right) \eea and \be \mathbb{A}_\mu =
e^{-\tga}\left(\partial_\mu+\frac{\partial}{\partial\vte^\mu}\tilde{A}-\vte_\mu\left(\eta-\vte^\nu\partial_\nu\bar{\Phi}-\left[\tilde{A},\bar{\Phi}\right]\right)\right)e^{\tga}
+ \te(\cdots) \ee One can check that the supersymmetry
transformations of the connection superfields reduce in components
to the known twisted transformation laws of the $\N$=2 super-
Yang--Mills theory in the Wess--Zumino gauge. This is obtained for
$\tilde{\ga}=\tilde{c}=0$ and redefining the supersymmetry
transformations by adding appropriated field-dependent super-gauge
transformations such that these fields are left invariant.


\subsection{Higher dimensions}

We now consider the $\N=1$, $d=10$ theory, which also encodes the
case $\N=2$, $d=8$. The constraints Eq.(\ref{constraints}) become
\be \label{constraints10D} \mathbf{\Upphi} =
\mathbb{L}_i=\bar{\mathbf{\Upphi}}_{ij} = 0,\quad
\boldsymbol{\upchi}_{ij}-\boldsymbol{\upchi}_{ji}+\frac{1}{3}\Omega_{ij}^{\phantom{ij}kl}\boldsymbol{\upchi}_{kl}=0.
\ee where $\Omega_{ijkl}$ is the octonionic eight-dimensional
$Spin(7)$-invariant $4$-form \cite{10DSYM}. Proceeding along the
same line as for the resolution of the constraints in four
dimensions, we get the gauge-fixed solution (once again we refer
the reader to \cite{TBP} for more details) \be \mathbb{A}_- = A_-
+ \te(\eta-\vte^i\partial_i A_-) \ee which gives the solution to
$\nabla_{d\vte}\mathbf{\Upgamma}+\mathbf{\Upgamma}^2=-\vert
d\vte\vert^2\mathbb{A}_-$ as $\mathbf{\Upgamma}_i=-\vte_i
\mathbb{A}_-$. Then, by introducing the functions $\tilde{A}$ and
$\tilde{A}_+$ of $\vte^i$ to parametrize $\mathbb{C}$, and by using
the constraints $\mathbf{\Upphi} = \mathbb{L}_i=0$ and the Bianchi
identities, one can write $\mathbb{A}_+$, $\mathbb{A}_i$ and $\boldsymbol{\upchi}_{ij}$ in terms of
$\mathbb{C}$ and $\mathbf{\Upgamma}_i$. Eventually, the
anti-selfdual constraint on $\boldsymbol{\upchi}_{[ij]}$ permits one
to completely determine the component field content of each
superfield. The expansion of $\tilde{A}$ and $\tilde{A}_+$ is in
fact \bea \tilde{A}\, &=& \vte^i A_i -
\frac{1}{2}\vte^i\vte^j\chi_{ij}-\frac{1}{3!}\vte^i\vte^j\vte^k\Omega_{ijk}^{\phantom{ijk}l}
F_{l-}+\cdots, \CR \tilde{A}_+ &=& A_+ -\vte^i \Psi_i
-\frac{1}{2}\vte^i\vte^j(F_{ij}+G_{ij}) + \cdots. \eea By
introducing the fields $ \tilde{c}$ and $ {\tga}$, one can
reinforce the super-gauge invariance and get the following
expression for the ten-dimensional superfield $\mathbb{C}$ \be
\mathbb{C} = \tilde{c} -
e^{-\tga}(\vte^i\partial_i+\tilde{A})e^{\tga} -\te
e^{-\tga}(\partial_+
+\tilde{A}_+)e^{\tga}-\te(\tilde{c}-e^{-\tga}(\vte^i\partial_i+\tilde{A})e^{\tga})^2.
\ee
 The supersymmetry transformation laws of the ten-dimensional super-Yang--Mills in components in the Wess--Zumino gauge \cite{10DSYM} are then recovered in an analogous way as in the four dimensional case.

\section{Action in superspace}

\subsection{The gauge invariant part}
We observe from the Bianchi identity
$\nabla\mathbf{\Upphi}+[\mathbb{C},\mathbf{\Upphi}]=0$ that
the gauge invariant function Tr\,$\mathbf{\Upphi}^2$ is $\te$ independent. Therefore, its components of highest order in $\vartheta^\mu$ can be used to
write the equivariant part of the action. The latter can be expressed as a full superspace integral of a Chern--Simons-like term
\be \label{actionEQ} \mathcal{S}_{EQ} = \int d^4\vte\, \trace
\mathbb{\mathbf{\Upphi}}^2= \int d^4\vte\,d\te\,\trace\Bigl(
\mathbb{C} \, \nabla \, \mathbb{C} + \frac{2}{3} \mathbb{C}^3
\Bigr) \ee
One can check that this action reproduces the known action for super-Yang--Mills in components. Notice that the superfield $\mathds{C}$ has a positive canonical dimension, which is an interesting point for its renormalization properties.

Unfortunately this formula does not generalize to higher dimensions. However, the one-loop invariant counter-terms involved in the eight-dimensional theory can be expressed as simple integrals over  superspace \be \int d^8 \vartheta \, \trace \Upphi^4 \hspace{10mm} \int d^8 \vartheta\, \trace \Upphi^2 \, \trace \Upphi^2 \ee

 The constraints can be covariantly implemented by the following superspace integral depending
on auxiliary Lagrange multipliers superfields
\begin{multline} \label{actionC}
\mathcal{S}_C = \int
d^4\vte\,d\te\,\trace\Bigl(
 \mathbb{B}^{(\mu\nu)}\bar{\mathbf{\Upphi}}_{\mu\nu} \, +{\scriptstyle \frac{1}{2}}
\bar{\mathbf{\Uppsi}}^{[\mu\nu]_+}\boldsymbol{\upchi}_{\mu\nu} +
\bar{\mathbb{K}}^\mu \mathbb{L}_\mu \Bigr)\\
= \int
d^4\vte\,d\te\,\trace\Bigl(
\mathbb{B}^{(\mu\nu)} \scal{\nabla_\mu\mathbf{\Upgamma}_\nu
+\mathbf{\Upgamma}_\mu \mathbf{\Upgamma}_\nu } \, +
{\scriptstyle \frac{1}{2}} \bar{\mathbf{\Uppsi}}^{[\mu\nu]_+}\scal{ \partial_\mu\mathbf{\Upgamma}_\nu
+\nabla_\mu\mathbb{A}_\nu+[\mathbb{A}_\mu,\mathbf{\Upgamma}_\nu]} \\
+\bar{\mathbb{K}}^\mu \scal{\nabla\mathbf{\Upgamma}_\mu +\nabla_\mu\mathbb{C}+\{\mathbf{\Upgamma}_\mu,\mathbb{C}\}+\mathbb{A}_\mu} \Bigr)
\end{multline}
where $\mathbb{B}^{(\mu\nu)}$ is symmetric traceless and
$\bar{\mathbf{\Uppsi}}^{[\mu\nu]_+}$ is antisymmetric selfdual.
The superfields $\bar{\mathbb{K}}_\mu$ and $\mathbb{A}_\mu$ can be
trivially integrated, giving rise to a simple substitution of
$\mathbb{A}_\mu$ by minus $ \nabla\mathbf{\Upgamma}_\mu
+\nabla_\mu\mathbb{C}+\{\mathbf{\Upgamma}_\mu,\mathbb{C}\}$. The resolution of the constraints is such that the formal integration over the auxiliary superfields $\mathbb{B}^{(\mu\nu)}$ and $\bar{\mathbf{\Uppsi}}^{[\mu\nu]_+}$ leads to the non-manifestly supersymmetric formulation of the theory in components, without introducing any determinant contribution in the path-integral. However, $\mathbb{B}^{(\mu\nu)}$ and $\bar{\mathbf{\Uppsi}}^{[\mu\nu]_+}$ admit a large class of zero
modes that must be considered in the manifestly supersymmetric superspace Feynman rules. They can be summarized by the following invariance of the action
\bea \label{zero} \delta^{\rm \scriptscriptstyle zero}
\mathbb{B}^{(\mu\nu)} &=& \hat{\nabla}_\sigma \scal{
\mathbf{\uplambda}^{(\sigma\mu\nu)} - \frac{1}{3} \hat{\nabla}
\mathbf{\upvarphi}^{\sigma(\mu,\nu)} } - \hat{\partial}_\sigma
\mathbf{\upvarphi}^{\sigma(\mu,\nu)} \CR \delta^{\rm
\scriptscriptstyle zero} \bar{\mathbf{\Uppsi}}^{[\mu\nu]_+} &=&
\hat{\nabla}_\sigma \mathbf{\upvarphi}^{[\mu\nu]_+, \sigma} \eea
where ${\uplambda}^{(\sigma\mu\nu)}$ is a superfield in the rank
three symmetric traceless representation and
${\upvarphi}^{[\mu\nu]_+, \sigma}$ is in the irreducible
representation defined by firstly taking the symmetric traceless
component in the two last indices and then projecting on the
antisymmetric selfdual component on the two first indices. These
gauge transformations are themselves invariant by a redefinition
of the superfields ${\uplambda}^{(\sigma\mu\nu)}$ and
${\upvarphi}^{[\mu\nu]_+, \sigma}$ by a gauge transformation
involving a superfield in the rank four symmetric traceless
representation and another one in the rank four irreducible
representation defined by firstly taking the symmetric traceless
component in the three last indices and then projecting on the
antisymmetric selfdual component on the two first indices. As a
matter of fact, the gauge-fixing of this gauge invariance requires
the introduction of an infinite set of ghosts including the ghosts
for ghosts, the ghosts for ghosts for ghosts and so on.

\subsection{The BRST symmetry and the gauge-fixing action in superspace.}
To fix the super-gauge invariance, one first introduces a Fadeev--Popov ghost superfield $\boldsymbol{\Upomega}$ and a BRST differential $\s$ that anticommutes with $\mathbf{\Updelta}$. As indicated by the super-gauge transformations (\ref{gauge_transf}) and their infinitesimal version, the BRST symmetry is defined as
\be
\s \mathcal{A} = - \mathbf{\Updelta}\boldsymbol{\Upomega}- [\mathcal{A},\boldsymbol{\Upomega}] , \hspace{7mm} \s \mathcal{F} = - [\boldsymbol{\Upomega},\mathcal{F}] , \hspace{7mm} \s \boldsymbol{\Upomega} = -\boldsymbol{\Upomega}^2 ,
\ee
One also needs a Fadeev--Popov antighost superfield
$\bar{\boldsymbol{\Upomega}}$ and its Lagrange multiplier superfield $\mathbb{B} $. In fact, the
BRST transformation laws of the super-connection, super-ghost and super-antighost follow from the following generalization of the horizontality equation Eq.(\ref{curvature}), which involves both the anti-BRST operator $\bar{\s}$ and the BRST operator $\s$
\be
(\mathbf{\Updelta}+ d \theta\, i_{d \vte
}+\s +\bar{\s})(\mathcal{A}+\boldsymbol{\Upomega} +\bar{\boldsymbol{\Upomega}})+(\mathcal{A}+\boldsymbol{\Upomega} +\bar{\boldsymbol{\Upomega}})^2
=
\mathcal{F},
\ee
This equation implies the degenerate equation $\s\bar{\boldsymbol{\Upomega}}+\bar{\s}\boldsymbol{\Upomega}+[\boldsymbol{\Upomega},\bar{\boldsymbol{\Upomega}}]=0$. It is solved by the introduction of the Lagrange multiplier superfield $\mathbb{B}$, so that one gets
\be
 \s \bar{\boldsymbol{\Upomega}} = \mathbb{B} , \hspace{7mm} \s \mathbb{B}=0, \hspace{7mm} \bar{\s}\boldsymbol{\Upomega}=-\mathbb{B}-[\boldsymbol{\Upomega},\bar{\boldsymbol{\Upomega}}]
\ee
A fully invariant gauge-fixing action can then be written as
\bea \label{actionGF} \mathcal{S}_{GF} = \s\bar{\s}\int
d^4\vte\,d\te\,\trace
\Bigl( \mathbb{A}_\mu \mathbb{A}^\mu \Bigr) = \s\int
d^4\vte\,d\te\,\trace\Bigl( \bar{\boldsymbol{\Upomega}} \, \partial^\mu
\mathbb{A}_\mu \Bigr)\CR = \int d^4\vte\,d\te\,\trace\Bigl(-\mathbb{B}\partial^\mu \mathbb{A}_\mu +\bar{\boldsymbol{\Upomega}}\partial^\mu
\hat{\partial}_\mu \boldsymbol{\Upomega}\Bigr)
\eea



One has also to write a gauge-fixing action for the action of
constraints. The gauge invariance (\ref{zero}) can be written in
terms of the BRST operator, thanks to the introduction of the
ghosts $\boldsymbol{\bar \Uppsi}^{\gra{1}{0} \mu\nu, \sigma}$ and
$\mathbb{B}^{\gra{1}{0} \mu\nu\sigma}$. As discussed in the
previous section, the BRST transformations are themselves subject
to a gauge invariance and one has to introduce an infinite tower of
ghosts for ghosts to correctly gauge-fix the theory. We define
the commuting ghosts $\boldsymbol{\bar \Uppsi}^{\gra{n}{0} \mu\nu,
\cdots }$ in the rank $n+2$ irreducible representation obtained by
applying the symmetric traceless projector on the $n+1$ last
indices and then the antisymmetric selfdual projector to the two
first indices, as well as the anticommuting ghost
$\mathbb{B}^{\gra{n}{0} \mu\nu\cdots}$ in the rank $n+2$ symmetric
traceless representation. The BRST transformations are the
following \bea
 \s \boldsymbol{\bar \Uppsi}^{\gra{n}{0} \mu\nu, \cdots } &=& \hat{\nabla}_\sigma \boldsymbol{\bar \Uppsi}^{\gra{n+1}{0} \mu\nu, \cdots \sigma} - [ \boldsymbol{\Upomega}, \boldsymbol{\bar \Uppsi}^{\gra{n}{0} \mu\nu, \cdots } ] \CR
 \s \mathbb{B}^{\gra{n}{0} \mu\nu\cdots} &=& \hat{\nabla}_\sigma \scal{ \mathbb{B}^{\gra{n+1 }{0} \mu\nu\cdots \sigma } + {\scriptstyle \frac{1}{n+3} }\hat{\nabla} \boldsymbol{\bar \Uppsi}^{\gra{n+1}{0}\sigma ( \mu, \nu  \cdots ) }} + \hat{\partial_\sigma} \boldsymbol{\bar \Uppsi}^{\gra{n+1}{0} \sigma (\mu, \nu \cdots )} - \{ \boldsymbol{\Upomega}, \mathbb{B}^{\gra{n}{0} \mu\nu\cdots} \} \CR
 \s \bar{\mathbb{K}}^\mu &=& \frac{1}{2} \hat{\nabla}_\sigma \hat{\nabla}_\nu \boldsymbol{\bar \Uppsi}^{\gra{1}{0} \mu\nu, \sigma } - \{ \boldsymbol{\Upomega},\mathbb{K}^\mu \}
 \eea
where $\boldsymbol{\bar \Uppsi}^{\gra{0}{0} \mu\nu}$ and $
\mathbb{B}^{\gra{0}{0} \mu\nu}$ are simply $\boldsymbol{\bar
\Uppsi}^{\mu\nu}$ and $ \mathbb{B}^{\mu\nu}$. The BRST operator is
nilpotent modulo the constraints, that is modulo the equations of
motion of the fields $\boldsymbol{\bar \Uppsi}^{\mu\nu}$, $
\mathbb{B}^{\mu\nu}$ and $\bar{\mathbb{K}}^\mu$. The
Batalin--Vilkovisky formalism permits one to solve this problem, by
introducing antifields as sources for the BRST
transformations.\footnote{However, we have not yet determined the
rank of the system, that is the maximal order at which the
antifields have to appear in the action.}

We have not yet worked out the gauge-fixing of this $BF$ system.
Even if it shares similarities with a standard bosonic $BF$ model,
the choice of gauge-functions cannot be defined by naively
replacing the space derivative of the bosonic case by the
anticommuting vector covariant derivative $\nabla_\mu$. It seems
that the free case can be worked out, by introducing transverse
projectors for the auxiliary fields, but more work is yet required
for a complete procedure. It will be described in the forthcoming
publication \cite{TBP}, as well as a practical way for doing
computations that takes into account the existence of the infinite
tower of ghosts in loops.

Despite our present ignorance of the gauge-fixing of the $BF$
system that enforces the covariant constraints, we thus propose as
a defining superspace action the following integral over the
twisted superspace \be \mathcal{S} = \mathcal{S}_{EQ} +
\mathcal{S}_C + \mathcal{S}_{GF} + \mathcal{S}_{CGF} \ee

The four-dimensional expressions (\ref{actionC}) and (\ref{actionGF}) of $ \mathcal{S}_C $ and $ \mathcal{S}_{GF} $
can be extended to eight and ten dimensions. It is not clear however if these expressions
are relevant in higher dimensions, where the introduction of a prepotential is required in order to write the equivariant part
of the action.
\section{Conclusion}

By using twisted variables, one can reexpress the $\N=2,d=4$
supersymmetry algebra in such a way that the pure super-Yang--Mills  theory is determined
by a subalgebra of the super-Poincar\'e algebra. We have seen the
existence of a corresponding twisted superspace, with coordinates
$(x^\mu,\theta,\vte^\mu)$. The result generalizes in higher dimensions. Quite interestingly, the constraints on the super-curvatures are such that they do not imply the equations
of motion. This general  property makes it plausible that one can obtain a
superspace path-integral formulation of maximally supersymmetric
theories.

Moreover, we have shown in this publication
that a twisted superspace path-integral formulation of the $\N = 2$
super-Yang--Mills theory does exist in four dimensions.
This theory is formulated as a Chern--Simons term  for the classical action plus   a $BF$
term for expressing the covariant constraints in superspace. Despite the fact that the gauge-fixing of the $BF$ part
requires the introduction of an infinite tower of ghosts and
ghosts for ghosts, we hope that it will
exhibit a general structure for a compact resumation of the ghost
contributions.  We have solved explicitly the constraints in component formalism and
verified that the theory reduces to the usual Yang--Mills theory
in components, after integration of the superspace longitudinal
components of the super-gauge fields and their corresponding
Faddeev--Popov ghosts.




Finally, it must be understood that the construction of a twisted superspace for the $\N=2$ supersymmetric theory is not an attempt for an alternative to its harmonic superspace formulation. Rather, it is a preliminary construction, as an example of a non-manifestly Lorentz invariant superspace-path-integral that can be generalized in ten dimensions, but must be completed within an harmonic superspace path-integral formulation for a complete description of the ten-dimensional super-Yang--Mills theory. Eventually, one expects the full Lorentz invariance to be recovered for the on-shell amplitudes.




\subsection*{Acknowledgments}

This work has been partially supported by the contract ANR (CNRS-USAR), \texttt{05-BLAN-0079-01}. A.~M. has been supported by the Swiss National Science Foundation, grant \texttt{PBSK2-119127}.


\end{document}